\documentclass[final,5p,times,twocolumn]{elsarticle}  

\usepackage{lineno}
\modulolinenumbers[5]

\journal{Journal of \LaTeX\ Templates}

\usepackage{amsmath}
\usepackage{amssymb}

\usepackage{graphicx}

\usepackage{hyperref}

\begin{document}

\begin{frontmatter}

\title{New Readout Scheme for Large Area Timing \& Position RPCs}


\author[mymainaddress]{João Saraiva\corref{mycorrespondingauthor}}

\cortext[mycorrespondingauthor]{Corresponding author}
\ead{joao.saraiva@coimbra.lip.pt}
\author[mymainaddress]{Alberto Blanco}
\address[mymainaddress]{LIP, Laboratory of Instrumentation and Experimental Particle Physics, Portugal}

\begin{abstract}
A new readout technique was developed with the primary aim of keeping the number of channels in the front-end electronics as low as possible when scaling up the sensitive area of a Resistive Plate Chamber (RPC). The readout method here presented significantly reduces the dependence between the detector area and the number of electronic channels, without substantial reduction of its performance: a 30 cm x 30 cm double stack multi-gap timing RPC was operated during weeks with cosmic rays, achieving a 2D spatial resolution well below 1 mm and time resolution lower than 100 ps, while its efficiency was kept above 98\%. 
\end{abstract}

\begin{keyword}
Gaseous Detectors, Resistive Plate Chambers, Readout Codification, Muon Scattering Tomography 
\end{keyword}

\end{frontmatter}

\section{Introduction}

Resistive Plate Chambers (RPCs) are deployed across diverse fields, for instance in High Energy Physics (HEP) for particle triggering and tracking over large areas, but also for Particle Identification (PID), if built to achieve high timing resolutions. In fact, with the capability of measuring simultaneously and accurately position and time, RPCs could actually perform PID without the need of additional detectors.
In the field of the Muon Scattering Tomography (MST), RPCs were first employed in 2012~\cite{Baesso-ref} to infer the presence of materials with high atomic number (Z). More recently, the MST technique was used anew with RPCs, showing the presence of several high Z materials after few minutes of acquisition~\cite{Saraiva-ref}. RPCs are indeed well suited for muographic techniques since they can be built at relatively low cost, covering large areas with high efficiency, spatial and time resolutions.
However, the front-end electronics (FEE), which constitute the major expense of the detector, can reach significant amounts when covering a surface well above 100 m$^2$, the area that must be instrumented to scan a shipping container.
A novel RPC readout codification was therefore designed and tested with the main purpose of addressing the aforementioned problem, partially decoupling the number of FEE channels from the sensitive area of the detector, while keeping high spatial and timing resolutions.

\section{Novel Readout Scheme} \label{sec-novelReadout}

With this new method, the strips of the readout electrodes are grouped in parallel by the Signal Merging Printed Circuit Board (SMPCB) as shown in Figure 1. In this way, each FEE channel reads out several strips in parallel, one per group, leading to a significant decrease of the number of preamplifiers, equivalent to the number of groups created by the SMPCB. 
However, the ambiguity that arises by grouping together strips of the thin-strip readout boards must be disentangled in order to determine in which group the signal was in fact induced. This is the role of the wide-strip readout electrode, which provides the 2D raw position of each event, allowing the impinged group to be identified in both directions.

In this setup, 120 thin strips were subdivided into 5 groups, resulting in the use of only 24 preamplifiers to read out all the strips. The reduction factor of the number of electronic channels could be even higher, by simply adding more groups of strips in parallel. In this way, a larger detector area would have been covered while keeping the number of FEE channels unchanged, decoupling both quantities, as pretended with this technique.

\begin{figure}[ht]
\centering
\includegraphics*[trim={7.7cm 6.5cm 7.6cm 3.cm}, width=0.6\linewidth]{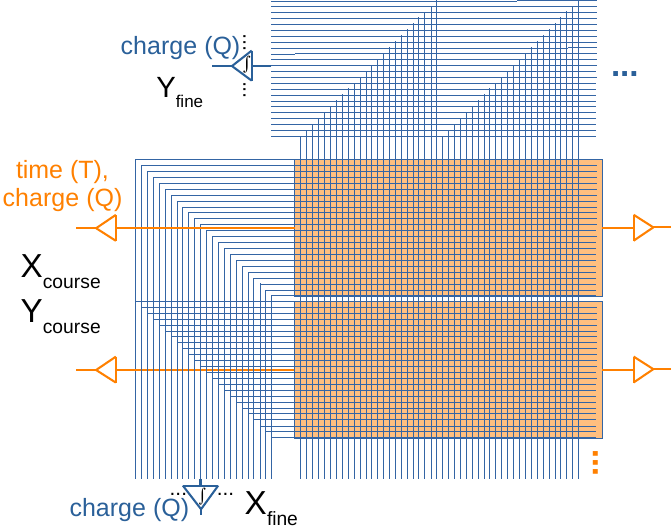}
\caption{Process of grouping in parallel the thin strips of two readout electrodes transversely oriented. A SMPCB is used for each direction. The wide strips used for disentanglement between groups are represented in orange.}
\label{fig-stripsSchematics}
\end{figure}

\section{Experimental Setup}

\begin{figure}[ht]
\centering
\includegraphics*[trim={0.cm 0.cm 0.cm 0.cm}, width=0.8\linewidth]{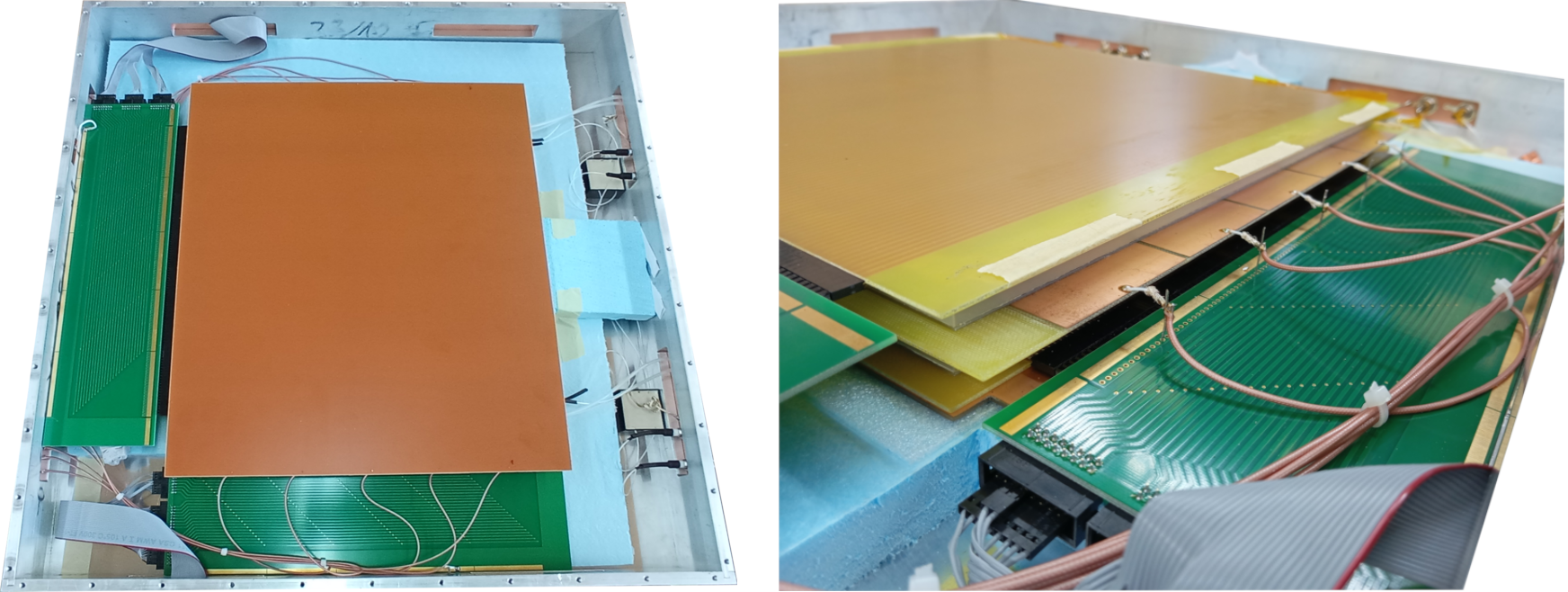}
\caption{(\textbf{left}) top view with the detector at the center and two SMPCBs grouping together the thin strips of the readout electrodes transversely oriented; (\textbf{right}) detailed view of the stack, from the bottom to the top: thin-strip PCB, RPC\_bottom, wide-strip PCB, RPC\_top, transversal thin-strip PCB.}
\label{fig-detector}
\end{figure}

Figure~\ref{fig-detector} shows the setup, composed of a stack of two multi-gap timing RPCs and three readout electrodes connected to the respective FEE. With an active area of 30 x 30 cm$^2$, the multi-gap RPCs have 6 gas gaps each, 300 $\mu$m wide, and resistive electrodes made of float glass, 1 mm thick ($\rho \approx$ 4x10$^{12}$~$\Omega$.cm at 25$^\circ$C). The detector was operated in open gas flow, with a gas mixture of 95.5\% of R-134a and 4.5\% of SF6.

Two types of pick-up electrodes were used in the setup:
\begin{itemize}
\item \textbf{wide-strip readout PCB}: constituted by 5 strips 5.9 cm wide and pitch of 6.1 cm; located between the RPCs, it is read out by fast preamps (based on HADES FEE~\cite{Belver-ref}) in both sides of the strips, providing simultaneously time and coarse 2D position of the interactions occurring in the sensitive volume of the detector, and making it possible to identify unequivocally, for each event, the group of thin strips where the signal induction took place;
\item \textbf{thin-strip readout PCB}: composed of 120 strips 1.54 mm wide and 2.54 mm pitch; two of them were used, one located at the bottom of the stack with the strips oriented in the same direction as the wide ones, and a second board at the top of the stack, with the strips transversely oriented with respect to the previous ones; both pick-up electrodes are connected to a SMPCB, reducing the number of channels from 120 to 24 (see Section~\ref{sec-novelReadout}); each of these channels (2x24) is then read out by slow electronics (custom-designed), integrating both components of the induced signals (electronic \& ionic), allowing, in this way, for submillimetric precision in positional measurements in both dimensions: X \& Y.
\end{itemize}

\section{Results}

\begin{figure}[ht]
\centering
\includegraphics*[trim={0.cm 0.cm 0.cm 0.cm}, width=0.8\linewidth]{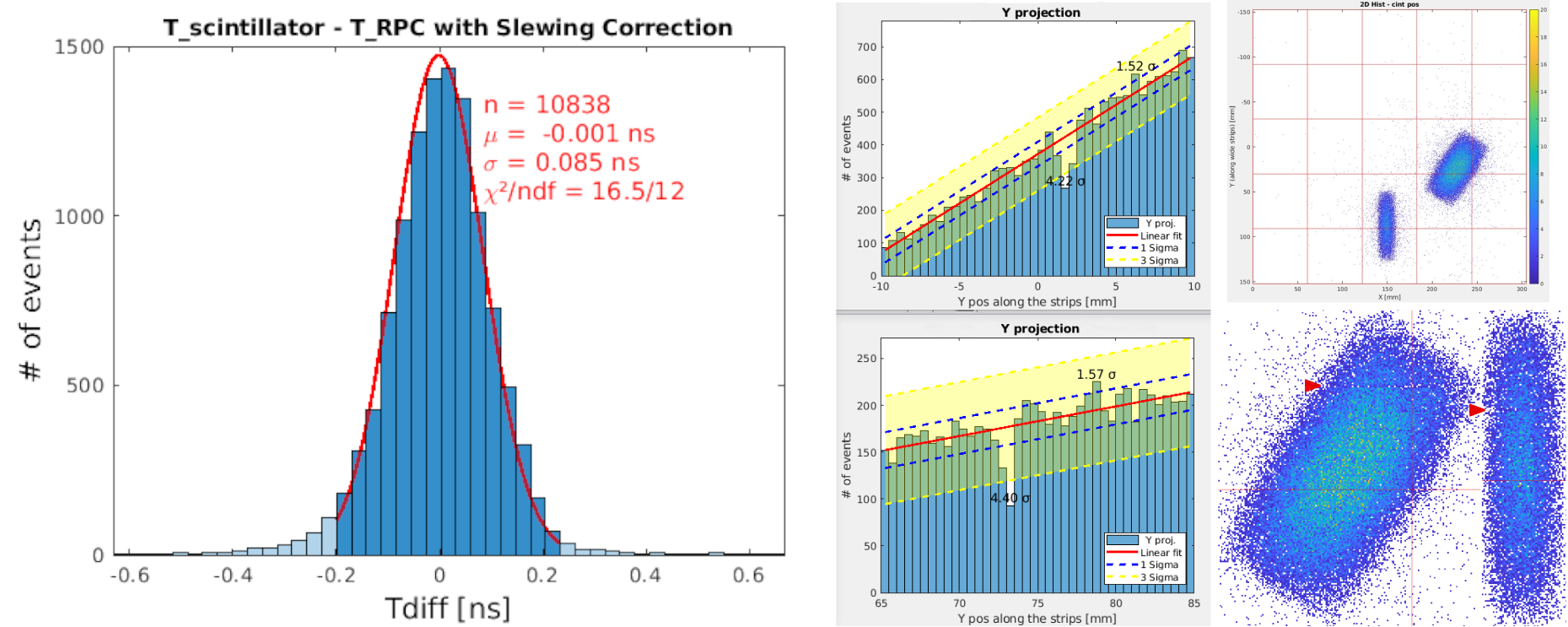}
\caption{(\textbf{left}) Time difference between RPCs and scintillators with a time resolution of 74 ps ($\sigma$) after: time walk correction and scintillator contribution removed; (\textbf{right}) right: projected shadow of the scintillators (the bottom image shows zoomed-in portions of the top one), left: respective projections on the Y axis, showing clearly the presence of the nylon monofilament spacers with a diameter of 300 $\mu$m (also visible in the zoomed-in image).}
\label{fig-results}
\end{figure}

The detector was operated during several weeks with cosmic rays, using a coincidence trigger generated externally by scintillators located above and below the RPCs.
Figure~\ref{fig-results} shows the achieved time and spatial resolutions when the reduced electric field was set to approximately 379 Td (around 2.75 kV/gap, 92 kV/cm) and the respective efficiency was slightly above 98\%.

\section{Application  - Muon Scattering Tomography}

With the proposed codification, it is possible to decouple two quantities usually directly correlated: the detector surface and the number of FEE channels needed to instrument the respective sensitive area. The decoupling is partial since, as seen previously, the channels connected to the wide strips have to increase proportionally with the number of groups of thin strips added in parallel. However, due to the large dimensions of the strips used for disentangling the groups, the increase of electronic channels achieved with this technique is by far lower than if each thin strip would have been connected to its own channel.

\begin{figure}[ht]
\centering
\includegraphics*[trim={0.cm 0.cm 0.cm 0.cm}, width=1.0\linewidth]{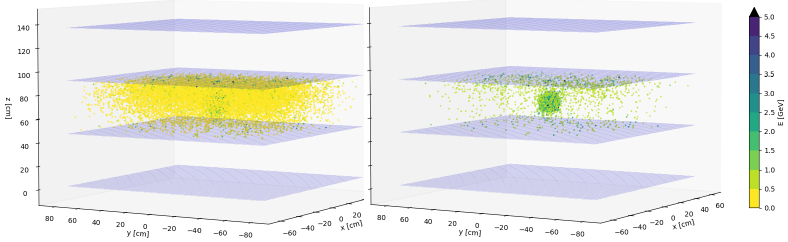}
\caption{Two step simulation using the FLUKA cosmic ray source and Earth atmospheric model~\cite{Fluka-ref}; geometry: MST tracking detector of four 2 m$^2$ RPCs, with a distance of 45 cm between planes and a tungsten block of 10x10x10 cm$^3$ at the center; applied restrictions: only scatters above 1.5$^\circ$ (\textbf{left}),
removing also muons below 500 MeV (\textbf{right})}
\label{fig-POCAs}
\end{figure}

Considering the above, and knowing that the FEE is the driving cost of gaseous detectors such as RPCs, the new readout scheme outlined in this document can be used in any application requiring large sensitive areas. Typical applications are linked to the HEP field, however another strong candidate is the MST, since:
\begin{itemize}
\item the area required for the MST of shipping containers is around 130 m$^2$, on par with the 141 m$^2$ covered by RPCs in the ALICE experiment at CERN~\cite{Alice-ref};	
\item the spatial resolution requirements for a muon tracking detector is dictated by the precision needed to measure the small scattering angles between the incoming and exiting muon trajectories; submillimetric spatial resolution allows both: (1) keeping the distance between detector planes, below and above the fiducial region, at few tens of centimeters (maximizing the detector acceptance and its compactness), (2) having the ability to detect small shifts of the muon direction, in the order of one degree (around 17 mrad).
\item a time resolution below 100 ps allows, via the Time Of Flight (TOF) technique, to reject low energy muons (few hundred of MeV) since they are highly scattered within the material budget of the detector, resulting in false events all around the fiducial region as shown in Figure~\ref{fig-POCAs};
\end{itemize}

\section{Conclusion}
A new readout scheme was developed to reduce the number of FEE channels (driving cost of the detector), and tested with a multi-gap timing RPC of 30 cm x 30 cm. With the presented codification, 24 + 24 charge sensitive preamps were used to read out 120 + 120 thin strips transversely oriented, providing submillimetric 2D positioning. An additional pick-up electrode was used with 5 + 5 current sensitive fast preamps to read out 5 large strips and resolve the ambiguity introduced by this technique. The experimental setup made it possible to achieve a 2D high spatial resolution ($<$ 1 mm), along with a time precision of 74 ps ($\sigma$) enabling PID, and an efficiency above 98\%. The next step includes using the same readout method with a large scale RPC (120 cm x 90 cm), without increasing the number of FEE channels connected to the thin strips, i.e. keeping the number of electronic channels independent of the sensitive area of the detector, as intended.

\section{Acknowledgements}
This work was supported by the 'Funda\c{c}\~{a}o para a Ci\^{e}ncia e Tecnologia, Portugal' (CERN/FIS-INS/0006/2021) and the European Union's Horizon 2020 Research and Innovation programme under Grant Agreement AIDAinnova n.$^\circ$ 101004761.


\end{document}